\documentclass[journal]{IEEEtran} 
\IEEEoverridecommandlockouts
\usepackage{cite}
\usepackage{amsmath,amssymb,amsfonts}
\usepackage{algorithmic}
\usepackage{bm}
\usepackage{tikz}
\usepackage{pgfplots}
\usetikzlibrary{calc}
\usetikzlibrary{arrows,shapes}
\usepackage{graphicx}
\usepackage{subfig}
\usepackage{textcomp}
\usepackage{xcolor}
\usepackage{soul}
\usepackage{amsthm}
\usepackage{siunitx}

\newtheorem{proposition}{Proposition}
\newtheorem{remark}{Remark}

\def\BibTeX{{\rm B\kern-.05em{\sc i\kern-.025em b}\kern-.08em
    T\kern-.1667em\lower.7ex\hbox{E}\kern-.125emX}}
\begin{document}
\title{EMF-Constrained Artificial Noise for Secrecy Rates with Stochastic Eavesdropper Channels
\thanks{This work was funded by the Ministry of Economic Affairs, Industry, Climate Action and Energy of the State of North Rhine-Westphalia, Germany under grant 005-2108-0021 (5G-Expo).}
}%
\author{\ 
    \IEEEauthorblockN{Stefan Roth}%
    , \IEEEauthorblockN{Aydin Sezgin}\\
    \IEEEauthorblockA{Ruhr University Bochum, Bochum, Germany}\\
    \IEEEauthorblockA{stefan.roth-k21@rub.de, aydin.sezgin@rub.de}
}%

\maketitle

\begin{abstract}
    An information-theoretic confidential communication is achievable if the eavesdropper has a degraded channel compared to the legitimate receiver. In wireless channels, beamforming and artificial noise can enable such confidentiality. However, only distribution knowledge of the eavesdropper channels can be assumed. Moreover, the transmission of artificial noise can lead to an increased electromagnetic field (EMF) exposure, which depends on the considered location and can thus also be seen as a random variable. Hence, we optimize the $\varepsilon$-outage secrecy rate under a $\delta$-outage exposure constraint in a setup, where the base station (BS) is communicating to a user equipment (UE), while a single-antenna eavesdropper with Rayleigh distributed channels is present. Therefore, we calculate the secrecy outage probability (SOP) in closed-form. Based on this, we convexify the optimization problem and optimize the $\varepsilon$-outage secrecy rate iteratively. Numerical results show that for a moderate exposure constraint, artificial noise from the BS has a relatively large impact due to beamforming, while for a strict exposure constraint artificial noise from the UE is more important.
\end{abstract}


\section{Introduction}

Wiretap channels enable an information-theoretic secure communication between different devices, if the eavesdropper has a degraded channel compared to the legitimate user \cite{6772207}. In this case, the secrecy rate is upper-bounded by the difference of the rates achieved by the legitimate user and the eavesdropper. While \cite{6772207} has focused on discrete, memoryless channels, the same statement has been proven for MIMO channels in \cite{5961840}. As the rates depend on the beamforming, the secrecy rate can be optimized by employing a narrow beamforming which steers the signal into the direction of the legitimate user and away from the eavesdropper. As additional optimization step, the concept of transmitting artificial noise towards the adversary to further degrade the eavesdropper's channel can be utilized as discussed in \cite{4543070}. In the process of optimizing the secrecy rate, various works (such as the works cited above) assume either the presence of perfect channel state information (CSI) or time-varying channels. The assumption of perfect CSI can be justified if the eavesdropper is part of the network. However, external eavesdroppers can also overhear the data transmissions passively and leave only marginal traces to the legitimate devices. In this case, only statistical CSI are available, such that the guarantee of a secrecy rate is often not possible. Therefore, knowledge on the channel distribution can be utilized to optimize the $\varepsilon$-outage secrecy rate \cite{5875883,10.1007/978-3-642-11526-4_1,6542749}, i.e., the secrecy rate of which outages are tolerated for a small portion $\varepsilon$ of the cases. Thereby, $\varepsilon$ can be referred to as the secrecy outage probability (SOP). The works \cite{5875883,10.1007/978-3-642-11526-4_1} have approximately characterized the $\varepsilon$-outage secrecy rate for the case that all devices are equipped with a single-antenna in closed-form for the case that no artificial noise is transmitted. In the work \cite{6542749}, a special beamformer design is considered, for which the SOP is calculated in closed-form for a multi-antenna eavesdropper and a receiver transmitting artificial noise in a full-duplex operation. In this work, we optimize the transmit beamforming and artificial noise transmission jointly, without restricting on a special structure of the beamforming design.

The transmission of artificial noise means that additional electromagnetic signals are broadcasted via the air, which leads to an increased electromagnetic field (EMF) exposure. When only considering a single transmitter, the exposure can be limited by introducing linear constraints to the beamformer optimization \cite{8886724}. However, the electromagnetic fields of the two transmitters can overlap randomly at different locations of the environment \cite{6565816}. To limit the EMF exposure in this case, we introduce a $\delta$-outage exposure constraint, i.e., a exposure constraint which must not be exceeded at more that a small portion $\delta$ of all locations (similar to the $\varepsilon$-outage secrecy rate).

\paragraph*{Contribution}
In this paper, we consider a setup with Rayleigh-distributed eavesdropper channels, where the transmitter and receiver are both equipped with two antennas, while the eavesdropper has a single antenna. Besides the transmit signal, the transmitter and receiver can both apply artificial noise to protect the transmission, such that the receiver utilizes one antenna for a full-duplex transmission. The contribution is as follows:
\begin{itemize}
    \item For the described setup, we formulate an \emph{optimization problem}, in which the the $\varepsilon$-outage secrecy rate is maximized under a $\delta$-outage exposure constraint.
    \item Moreover, we derive exact \emph{closed-form expressions of the SOP and the probability of exceeding a certain exposure level} for the considered setup.
    \item Afterwards, we formulate \emph{convex approximations} of these expressions and optimize the $\varepsilon$-outage secrecy rate iteratively.
\end{itemize}

\paragraph*{Notation} Scalars, vectors and matrices are denoted as regular, bold lower-case and bold-upper-case letters, respectively. $\|\bm{A}\|_F$ is the Froebenius norm, while $\bm{A}\succeq\bm{0}$ indicates that $\bm{A}$ is positive semi-definite. $\mathcal{CN}(\bm{a},\bm{B})$ refers to a circularly-symmetric Gaussian distribution with mean $\bm{a}$ and covariance $\bm{B}$, while $Q(\cdot,\cdot)$ is the regularized upper incomplete gamma function. $\mathrm{Prob}(p)$ is the probability that the random Boolean variable $p$ is true.

\section{System Model}
\begin{figure}
    \centering
    \includegraphics{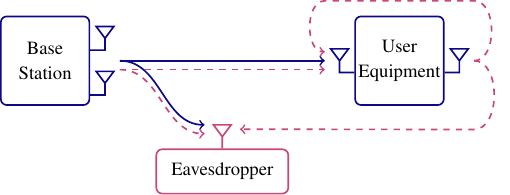}
    \caption{The base station is transmitting data to the user (solid lines). To protect the data transmission against eavesdropping, both devices can broadcast artificial noise (dashed lines), which also causes self-interference as a side-effect.}
    \label{fig:motivation}
\end{figure}

We consider a downlink setup as shown in \figurename~\ref{fig:motivation}, where the base station (BS) is equipped with $N=2$ antennas, while the user equipment (UE) has $\bar{N}_{\mathrm{T}}=1$ transmit and $\bar{N}_{\mathrm{R}}=1$ receive antenna. This means that the UE operates in a full-duplex mode. The eavesdropper is equipped with a single antenna. The BS regularly transmits symbol vectors $\bm{s}$ to the UE from a Gaussian codebook, i.e., distributed as $\bm{s}\sim\mathcal{CN}(\bm{0},\bm{I}_{N})$. To protect the message from being overheard at an eavesdropper, the BS and the UE can both sends artificial noise symbol vectors $\bm{s}_{\mathrm{n}}$ and $\bm{\bar{s}}_{\mathrm{n}}$ simultaneously, which are also drawn from a Gaussian codebook, and thus distributed as $\bm{s}_{\mathrm{n}}\sim\mathcal{CN}(\bm{0},\bm{I}_{N})$ and $\bm{\bar{s}}_{\mathrm{n}}\sim\mathcal{CN}(\bm{0},\bm{I}_{\bar{N}_{\mathrm{T}}})$. The BS applies the beamforming matrices $\bm{W}$ and $\bm{W}_{\mathrm{n}}$ of size $N\times N$, while the UE applies the beamforming matrix $\bm{\bar{W}}_{\mathrm{n}}$ of size $\bar{N}_{\mathrm{T}}\times \bar{N}_{\mathrm{T}}$. Hence, the transmit signals of the BS and the UE are
\begin{align}
    \bm{x}&=\bm{W}\bm{s}+\bm{W}_{\mathrm{n}}\bm{s}_{\mathrm{n}}\\
    \bm{\bar{x}}&=\bm{\bar{W}}_{\mathrm{n}}\bm{\bar{s}}_{\mathrm{n}}.
\end{align} 
The transmit power of the two transmissions are limited by $P^{\mathrm{max}}$ and $\bar{P}^{\mathrm{max}}$.

We consider Rayleigh fading channels between the devices. The channel between the BS and the UE is referred to as $\bm{H}_{\mathrm{U}}$, while the channel between the BS and the eavesdropper is $\bm{h}_{\mathrm{E}}$ and the channel between the UE and the eavesdropper is $\bm{\bar{h}}_{\mathrm{E}}$. Due to the full-duplex operation, self-interference occurs at the user equipment; the according channel is described as $\bm{\bar{H}}_{\mathrm{U}}$. Hence, the received signals at the UE and the eavesdropper are
\begin{align}
    \bm{y}_{\mathrm{U}}&=\bm{H}_{\mathrm{U}}\bm{x}+\bm{\bar{H}}_{\mathrm{U}}\bm{\bar{x}}+\bm{v}_{\mathrm{U}},\\
    y_{\mathrm{E}}&=\bm{h}_{\mathrm{E}}^H\bm{x}+\bm{\bar{h}}_{\mathrm{E}}^H\bm{\bar{x}}+v_{\mathrm{E}}.
\end{align}
Thereby, AWGN is represented by the Gaussian-distributed variables $\bm{v}_{\mathrm{U}}\sim\mathcal{CN}\left(\bm{0},\bm{V}_{\mathrm{U}}\right)$ and $v_{\mathrm{E}}\sim\left(\bm{0},V_{\mathrm{E}}\right)$. 

For the channel $\bm{H}_{\mathrm{U}}$, we assume the knowledge of perfect CSI. The self-interference channel $\bm{\bar{H}}_{\mathrm{U}}$ is modelled as a compound channel (see \cite{7071563}). 
This means that an estimate of $\bm{\bar{H}}_{\mathrm{U}}$ is present at the UE, which contains an uncertainty $\bm{\bar{H}'}_{\mathrm{U}}$ bounded by $\mathrm{trace}\{\bm{\bar{H}'}_{\mathrm{U}}\left(\bm{\bar{H}'}_{\mathrm{U}}\right)^H\}\leq \bar{G}_{\mathrm{U}}^{\mathrm{max}}$. Thereby, $\bar{G}_{\mathrm{U}}^{\mathrm{max}}$ is the upper-bound of the channel gain. To remove the impact of the self-interference from the received signal, the UE employs successive interference cancellation (SIC), after which only the uncertain part remains. 
Among the possible set of residuals, we consider the worst-case channel to obtain the rate which can be achieved for all channel realizations. 
Hence, in the considered worst-case, the rate of the message $s$ becomes
\begin{align}
    R_{\mathrm{U}}&=\log_2\bigg(\mathrm{det}\bigg(\bm{I}_{\bar{N}_{\mathrm{R}}}+\bm{H}_{\mathrm{U}}\bm{W}\bm{W}^H\bm{H}_{\mathrm{U}}^H\nonumber\\&\hspace{0.5cm}\times\big(\bar{G}_{\mathrm{U}}^{\mathrm{max}}\mathrm{trace}\left\{\bm{\bar{W}}_{\mathrm{n}}\bm{\bar{W}}_{\mathrm{n}}^H\right\}\bm{I}_{\bar{N}_{\mathrm{R}}}\nonumber\\&\hspace{1cm}+\bm{H}_{\mathrm{U}}\bm{W}_{\mathrm{n}}\bm{W}_{\mathrm{n}}^H\bm{H}_{\mathrm{U}}^H+\bm{V}_{\mathrm{U}}\big)^{-1}\bigg)\bigg).\label{eq:R_UB}
\end{align}
The eavesdropper is assumed to have full CSI of its channels, but unable to decode any part of the artificial noise. Thus, it decodes the message $s$ with a rate of
\begin{align}
    R_{\mathrm{E}}&=\log_2\left(1+\frac{\bm{h}_{\mathrm{E}}^H\bm{W}\bm{W}^H\bm{h}_{\mathrm{E}}}{\bm{\bar{h}}_{\mathrm{E}}^H\bm{\bar{W}}_{\mathrm{n}}\bm{\bar{W}}_{\mathrm{n}}^H\bm{\bar{h}}_{\mathrm{E}} +\bm{h}_{\mathrm{E}}^H\bm{W}_{\mathrm{n}}\bm{W}_{\mathrm{n}}^H\bm{h}_{\mathrm{E}}+V_{\mathrm{E}}}\right).
\end{align}
Using these two rate expressions, the secrecy rate is \cite{6772207,7270404}
\begin{align}
    R=\mathrm{max}\left(R_{\mathrm{U}}-R_{\mathrm{E}},0\right).
\end{align}
In practice, the base station and the UE both typically only know the distributions of $\bm{H}_{\mathrm{E}}$ and $\bm{H}_{\mathrm{U}}$, but have no knowledge about the actual channel realizations. Due to Rayleigh fading, the distributions are
\begin{align}
    \bm{h}_{\mathrm{E}}&\sim\mathcal{CN}\left(\bm{0},\bm{G}_{\mathrm{E}}\right),&
    \bm{\bar{h}}_{\mathrm{E}}&\sim\mathcal{CN}\left(\bm{0},\bm{\bar{G}}_{\mathrm{E}}\right),\nonumber
\end{align}
where $\bm{G}_{\mathrm{E}}$ and $\bm{\bar{G}}_{\mathrm{E}}$ are the covariance matrices of the eavesdropper channels.
Due to the randomness of the channel vectors, $R_{\mathrm{E}}$ is also a (potentially unbounded) random variable. Thus, we consider the $\varepsilon$-outage secrecy rate (see \cite{6542749}), i.e., a secrecy rate which is fallen below only for a portion $\varepsilon$ of all eavesdropper channel realizations, i.e.,
\begin{align}
    &R_{\varepsilon}=\max\left(R_{\mathrm{U}}-R_{\mathrm{E}}^{\mathrm{max}},0\right),\\&\mathrm{Prob}(R_{\mathrm{E}}\leq R_{\mathrm{E}}^{\mathrm{max}})\geq 1-\varepsilon.
\end{align}

Within this setup, the EMF exposure obtained at an arbitrary location with channels $\bm{h}_{\mathrm{D}}$ and $\bm{\bar{h}}_{\mathrm{D}}$ is specified by
\begin{align}
    P_{\mathrm{D}}=\bm{h}_{\mathrm{D}}^H\left(\bm{W}\bm{W}^H+\bm{W}_{\mathrm{n}}\bm{W}_{\mathrm{n}}^H\right)\bm{h}_{\mathrm{D}}+\bm{\bar{h}}_{\mathrm{D}}^H\bm{\bar{W}}_{\mathrm{n}}\bm{\bar{W}}_{\mathrm{n}}^H\bm{\bar{h}}_{\mathrm{D}}.
\end{align}
Here, the channels $\bm{h}_{\mathrm{D}}$ and $\bm{\bar{h}}_{\mathrm{D}}$ are also modelled as Rayleigh-fading channels, such that they can be described as
\begin{align}
    \bm{h}_{\mathrm{D}}&\sim\mathcal{CN}\left(\bm{0},\bm{G}_{\mathrm{D}}\right),&
    \bm{\bar{h}}_{\mathrm{D}}&\sim\mathcal{CN}\left(\bm{0},\bm{\bar{G}}_{\mathrm{D}}\right).\nonumber
\end{align}
Thereby, $\bm{G}_{\mathrm{D}}$ and $\bm{\bar{G}}_{\mathrm{D}}$ are the covariance matrices of the corresponding channels. From the above distributions, $P_{\mathrm{D}}$ is a stochastic variable. Hence, the $\delta$-outage exposure constraint can be formalized to ensure that $P_{\mathrm{D}}$ does not exceed $P_{\mathrm{D}}^{\mathrm{max}}$ with the exception of a small probability $\delta$, i.e.,
\begin{align}
    \mathrm{Prob}\left(P_{\mathrm{D}}\leq P_{\mathrm{D}}^{\mathrm{max}}\right)\geq 1-\delta.
\end{align}

\section{Optimization Problem}
We aim at maximizing the $\varepsilon$-outage secrecy rate, while the $\delta$-outage exposure constraint and additional power constraints are met. The optimization problem can be formulated as
\begin{subequations}\label{eq:prob-1}
\begin{align}
    \underset{\substack{\bm{W},\bm{W}_n,\\\bm{\bar{W}}_n,R_{\mathrm{E}}^{\mathrm{max}}}}{\mathrm{maximize}}\hspace{0.2cm} & \max\left(R_{\mathrm{U}}-R_{\mathrm{E}}^{\mathrm{max}},0\right)\tag{\ref{eq:prob-1}}\\
    \mathrm{subject\ to}\hspace{0.2cm}&\mathrm{Prob}(R_{\mathrm{E}}\leq R_{\mathrm{E}}^{\mathrm{max}})\geq 1-\varepsilon\label{eq:prob-1:eve}\\
    &\mathrm{Prob}\left(P_{\mathrm{D}}\leq P_{\mathrm{D}}^{\mathrm{max}}\right)\geq 1-\delta\label{eq:prob-1:emf}\\
    &\mathrm{trace}\left\{\bm{W}\bm{W}^H\right\}+\mathrm{trace}\left\{\bm{W}_{\mathrm{n}}\bm{W}_{\mathrm{n}}^H\right\}\leq P^{\mathrm{max}}\label{eq:prob-1:PB}\\
    &\mathrm{trace}\left\{\bm{\bar{W}}_{\mathrm{n}}\bm{\bar{W}}_{\mathrm{n}}^H\right\}\leq \bar{P}^{\mathrm{max}}\label{eq:prob-1:PU}
\end{align}
\end{subequations}
While the constraints 
\eqref{eq:prob-1:PB} and \eqref{eq:prob-1:PU} are convex, this does not hold true for the objective \eqref{eq:prob-1} and the constraints \eqref{eq:prob-1:eve} and \eqref{eq:prob-1:emf}. In the following subsections, we will derive convex expressions for this three terms, and then formulate a convex version of the optimization problem.

\subsection{Target Rate}
The objective contains \eqref{eq:R_UB}, which is non-convex. To convexify this term, we first note that the precoding matrices appear only as squares within the optimization problem. Hence, we replace the precoders by their covariance matrices, i.e., $\bm{Q}=\bm{W}\bm{W}^H$, $\bm{Q}_{\mathrm{n}}=\bm{W}_{\mathrm{n}}\bm{W}_{\mathrm{n}}^H$ and $\bm{\bar{Q}}_{\mathrm{n}}=\bm{\bar{W}}_{\mathrm{n}}\bm{\bar{W}}_{\mathrm{n}}^H$. Note that this step requires to ensure that the covariance matrices are positive semidefinite and hermitian. Next, we create a convex lower-bound of \eqref{eq:R_UB} by applying Fenchel's inequality from \cite{8581021,citeulike:163662}, i.e., we lower-bound the rate as $R_{\mathrm{U}}\geq \Tilde{R}_{\mathrm{U}}$, where
\begin{align}
    \Tilde{R}_{\mathrm{U}}&=\log_2\left(\mathrm{det}\left(\bm{A}\right)\right)-\log_2\left(\mathrm{det}\left(\bm{B}^\star\right)\right)\nonumber\\&\hspace{1cm}-\mathrm{trace}\left\{\left(\bm{B}^\star\right)^{-1}\bm{B}\right\}+\bar{N}_{\mathrm{R}},\label{eq:objective:reformulated}
\end{align}
and
\begin{align}
    \bm{A}&=\bm{H}_{\mathrm{U}}\left(\bm{Q}+\bm{Q}_{\mathrm{n}}\right)\bm{H}_{\mathrm{U}}^H+\bar{G}_{\mathrm{U}}^{\mathrm{max}}\mathrm{trace}\left\{\bm{\bar{Q}}_{\mathrm{n}}\right\}\bm{I}_{\bar{N}_{\mathrm{R}}}+\bm{V}_{\mathrm{U}},\\
    \bm{B}&=\bar{G}_{\mathrm{U}}^{\mathrm{max}}\mathrm{trace}\left\{\bm{\bar{Q}}_{\mathrm{n}}\right\}\bm{I}_{\bar{N}_{\mathrm{R}}}+\bm{H}_{\mathrm{U}}\bm{Q}_{\mathrm{n}}\bm{H}_{\mathrm{U}}^H+\bm{V}_{\mathrm{U}}.
\end{align}
Moreover, $\bm{B}^\star$ is an operation point in $\bm{B}$. When employing $\Tilde{R}_{\mathrm{U}}$ in the objective instead of $R_{\mathrm{U}}$, the optimization problem becomes an iterative problem, where the value of $\bm{B}^\star$ is updated in every iteration (similar to the algorithm in \cite{8581021}). After optimization, the original precoder can be recovered via eigenvalue decomposition or singular value decomposition.

\subsection{Eavesdropper Rate}
To tackle the eavesdropper term \eqref{eq:prob-1:eve}, we first aim at obtaining a closed-form expression of $\mathrm{Prob}(R_{\mathrm{E}}\leq R_{\mathrm{E}}^{\mathrm{max}})$. Therefore, we define $\bm{L}_{\mathrm{E}}$ and $\bm{\bar{L}}_{\mathrm{E}}$ via Cholesky decomposition such that $\bm{G}_{\mathrm{E}}=2\bm{L}_{\mathrm{E}}^H\bm{L}_{\mathrm{E}}$ and $\bm{\bar{G}}_{\mathrm{E}}=2\bm{\bar{L}}_{\mathrm{E}}^H\bm{\bar{L}}_{\mathrm{E}}$. $\bm{\Lambda}_{\mathrm{E}}$ and $\bm{\bar{\Lambda}}_{\mathrm{E}}$ are the diagonal eigenvalue matrices of $2\bm{L}_{\mathrm{E}}\left(\bm{W}\bm{W}^H-\left(2^{R_{\mathrm{E}}^{\mathrm{max}}}-1\right)\bm{W}_{\mathrm{n}}\bm{W}_{\mathrm{n}}^H\right)\bm{L}_{\mathrm{E}}^H$ and $-2\left(2^{R_{\mathrm{E}}^{\mathrm{max}}}-1\right)\bm{\bar{L}}_{\mathrm{E}}\bm{\bar{W}}_{\mathrm{n}}\bm{\bar{W}}_{\mathrm{n}}^H\bm{\bar{L}}_{\mathrm{E}}^H$, respectively. Furthermore, we have $z_{\mathrm{E}}=\left(2^{R_{\mathrm{E}}^{\mathrm{max}}}-1\right)V_{\mathrm{E}}$. Based on these definitions, the closed-form expression of $\mathrm{Prob}(R_{\mathrm{E}}\leq R_{\mathrm{E}}^{\mathrm{max}})$ is provided in Proposition~\ref{prop:eve:reformulated:general}.
\begin{proposition}\label{prop:eve:reformulated:general}
    When grouping equal eigenvalues of $\mathrm{diag}(\bm{\Lambda}_{\mathrm{E}},\bm{\bar{\Lambda}}_{\mathrm{E}})$ together and defining $\lambda_k$ as the distinct eigenvalues and $m_k$ as the corresponding quantity, the term $\mathrm{Prob}(R_{\mathrm{E}}\leq R_{\mathrm{E}}^{\mathrm{max}})$ equals for the considered parameters
    \begin{align}
        \mathrm{Prob}(R_{\mathrm{E}}\leq R_{\mathrm{E}}^{\mathrm{max}})&=1-\sum_{\{k|\lambda_k > 0\}}\mathrm{exp}\left(-\Upsilon_{k}\right)\nonumber\\&\hspace{-1.5cm}\times Q\left(m_k,\frac{z_{\mathrm{E}}}{\lambda_k}-\Upsilon_{k}\right)\prod_{j\neq k}\left(\frac{\lambda_k}{\lambda_k-\lambda_j}\right)^{m_j},\label{eq:prob:modified}
\end{align}
where $\Upsilon_{k}=\sum_{j\neq k}m_j\lambda_j\left(\lambda_k-\lambda_j\right)^{-1}$. 
\end{proposition}
\begin{proof}
    Due to lack of space, the full details of the proof are not provided here. Rather the major steps are described shortly. The term $\mathrm{Prob}(R_{\mathrm{E}}\leq R_{\mathrm{E}}^{\mathrm{max}})$ is equal to $1-\mathrm{Prob}\left(\frac{1}{2}\bm{r}_{\mathrm{E}}^H\bm{A}_{\mathrm{E}}\bm{r}_{\mathrm{E}}\geq z_{\mathrm{E}}\right)$,
    where $\bm{r}_{\mathrm{E}}\sim\mathcal{CN}\left(\bm{0},\bm{I}_{N+\bar{N}_{\mathrm{T}}}\right)$ and $\bm{A}_{\mathrm{E}}=\mathrm{diag}\left(\bm{\Lambda}_{\mathrm{E}},\ \bm{\bar{\Lambda}}_{\mathrm{E}}\right)$.
    A closed form solution for such an expression has been provided by \cite{490565}. The solution is adapted for the case of duplicated eigenvalues through the rule of de L'Hospital. Then, the series is extended such that the result is exact for larger $m_k$ if all eigenvalues are equal, which leads to \eqref{eq:prob:modified}. 
\end{proof}

To obtain a concave approximation of \eqref{eq:prob:modified}, we again replace $\bm{W}\bm{W}^H$, $\bm{W}_{\mathrm{n}}\bm{W}_{\mathrm{n}}^H$ and $\bm{\bar{W}}_{\mathrm{n}}\bm{\bar{W}}_{\mathrm{n}}^H$ by $\bm{Q}$, $\bm{Q}_{\mathrm{n}}$ and $\bm{\bar{Q}}_{\mathrm{n}}$. Within the operation point $\left(\bm{Q}^\star,\bm{Q}_{\mathrm{n}}^\star,\bm{\bar{Q}}_{\mathrm{n}}^\star,R_{\mathrm{E}}^{\mathrm{max},\star}\right)$, the first-order Taylor approximation becomes
\begin{align}
    T_{\mathrm{E}}&=
    C_{\mathrm{E},0}^\star+\mathrm{trace}\left\{\left(\bm{Q}-\bm{Q}^\star\right)\bm{C}_{\mathrm{E}}\right\}\nonumber\\&\hspace{0.1cm}+\mathrm{trace}\left\{\left(\bm{Q}_{\mathrm{n}}-\bm{Q}_{\mathrm{n}}^\star\right)\bm{C}_{\mathrm{E},\mathrm{n}}\right\}+\mathrm{trace}\left\{\left(\bm{\bar{Q}}_{\mathrm{n}}-\bm{\bar{Q}}_{\mathrm{n}}^\star\right)\bm{\bar{C}}_{\mathrm{E},\mathrm{n}}\right\}\nonumber\\&\hspace{0.1cm}+\left(R_{\mathrm{E}}^{\mathrm{max}}-R_{\mathrm{E}}^{\mathrm{max,\star}}\right)C_{\mathrm{E},\mathrm{R}}.\label{eq:eve:general:convex}
\end{align}
The variables in \eqref{eq:eve:general:convex} are provided in appendix~\ref{appendix:taylor_parameters}.
\subsection{Exposure Constraint}
The exposure constraint can be convexified very similar to the eavesdropper rate. Therefore, first a closed-form expression of $\mathrm{Prob}(P_{\mathrm{D}}\leq P_{\mathrm{D}}^{\mathrm{max}})$ is needed. To obtain such, we first employ any decomposition that fulfills $\bm{G}_{\mathrm{D}}=2\bm{L}_{\mathrm{D}}^H\bm{L}_{\mathrm{D}}$ and $\bm{\bar{G}}_{\mathrm{D}}=2\bm{\bar{L}}_{\mathrm{D}}^H\bm{\bar{L}}_{\mathrm{D}}$ to obtain $\bm{L}_{\mathrm{D}}$ and $\bm{\bar{L}}_{\mathrm{D}}$. Further, $\bm{\Lambda}_{\mathrm{D}}$ and $\bm{\bar{\Lambda}}_{\mathrm{D}}$ are the diagonal eigenvalue matrices of $2\bm{L}_{\mathrm{D}}\left(\bm{W}\bm{W}^H+\bm{W}_{\mathrm{n}}\bm{W}_{\mathrm{n}}^H\right)\bm{L}_{\mathrm{D}}^H$ and $2\bm{\bar{L}}_{\mathrm{D}}\bm{\bar{W}}_{\mathrm{n}}\bm{\bar{W}}_{\mathrm{n}}^H\bm{\bar{L}}_{\mathrm{D}}^H$, respectively. When further defining $z_{\mathrm{D}}=P_{\mathrm{D}}^{\mathrm{max}}$, the term $\mathrm{Prob}(P_{\mathrm{D}}\leq P_{\mathrm{D}}^{\mathrm{max}})$ can be obtained similar to the result of Proposition~\ref{prop:eve:reformulated:general}. 

Note that \eqref{eq:prob:modified} is still non-convex. A convex approximation can be obtained by calculating a first-order Taylor series around the operating point, which equals
\begin{align}
    T_{\mathrm{D}}&=C_{\mathrm{D},0}^\star+\mathrm{trace}\left\{\left(\bm{Q}+\bm{Q}_{\mathrm{n}}-\bm{Q}^\star-\bm{Q}_{\mathrm{n}}^\star\right)\bm{C}_{\mathrm{D}}\right\}\nonumber\\&\hspace{1cm}+\mathrm{trace}\left\{\left(\bm{\bar{Q}}_{\mathrm{n}}-\bm{\bar{Q}}_{\mathrm{n}}^\star\right)\bm{\bar{C}}_{\mathrm{D}}\right\}.\label{eq:emf:general:convex}
\end{align}
The variables $\bm{C}_{\mathrm{D}}$ and $\bm{\bar{C}}_{\mathrm{D}}$ are again provided in appendix~\ref{appendix:taylor_parameters}.

\subsection{Convex Problem}
To summarize the previous subsections, a convex version of problem \eqref{eq:prob-1} can be obtained by employing the covariance matrices $\bm{Q}$, $\bm{Q}_{\mathrm{n}}$ and $\bm{\bar{Q}}_{\mathrm{n}}$ together with \eqref{eq:objective:reformulated}, \eqref{eq:eve:general:convex} and \eqref{eq:emf:general:convex}. 
To ensure convergence, we additionally penalize differences between the optimization variables $R_{\mathrm{E}}^{\mathrm{max}}$, $\bm{Q}$, $\bm{Q}_{\mathrm{n}}$ and $\bm{\bar{Q}}_{\mathrm{n}}$ and the according operation points $R_{\mathrm{E}}^{\mathrm{max},\star}$, $\bm{Q}^\star$, $\bm{Q}_{\mathrm{n}}^\star$ and $\bm{\bar{Q}}_{\mathrm{n}}^\star$ in the objective.
\begin{subequations}\label{eq:prob-2}
\begin{align}
    \underset{\bm{Q},\bm{Q}_{\mathrm{n}},\bm{\bar{Q}}_{\mathrm{n}},R_{\mathrm{E}}^{\mathrm{max}}}{\mathrm{maximize}}\hspace{0.2cm} & \Tilde{R}_{\mathrm{U}}-R_{\mathrm{E}}^{\mathrm{max}}-\Gamma_\mathrm{R}\left|R_{\mathrm{E}}^{\mathrm{max}}-R_{\mathrm{E}}^{\mathrm{max},\star}\right|^2\hspace{-1.9cm}\nonumber\\&\hspace{-1.5cm}-\Gamma\left\|\bm{Q}-\bm{Q}^\star\right\|_{F}^2-\Gamma_{\mathrm{n}}\left\|\bm{Q}_{\mathrm{n}}-\bm{Q}_{\mathrm{n}}^\star\right\|_{F}^2-\bar{\Gamma}_{\mathrm{n}}\left\|\bm{\bar{Q}}_{\mathrm{n}}-\bm{\bar{Q}}_{\mathrm{n}}^\star\right\|_{F}^2\tag{\ref{eq:prob-2}}\\
    \mathrm{subject\ to}\hspace{0.2cm}&T_{\mathrm{E}}\geq1-\varepsilon\label{eq:prob-2:eve}\\
    &T_{\mathrm{D}}\geq1-\delta\label{eq:prob-2:emf}\\
    &\mathrm{trace}\left\{\bm{Q}\right\}+\mathrm{trace}\left\{\bm{Q}_{\mathrm{n}}\right\}\leq P^{\mathrm{max}}\hspace{-1cm}\label{eq:prob-2:PB}\\
    &\mathrm{trace}\left\{\bm{\bar{Q}}_{\mathrm{n}}\right\}\leq \bar{P}^{\mathrm{max}}\label{eq:prob-2:PU}\\
    &\bm{Q}\succeq \bm{0},\ \bm{Q}_{\mathrm{n}}\succeq \bm{0},\ \bm{\bar{Q}}_{\mathrm{n}}\succeq \bm{0}
\end{align}
\end{subequations}
This problem can be solved iteratively via semidefinite programming by a convex solver, as done here via CVX \cite{cvx}, until convergence. The variables $\Gamma_{\mathrm{R}}$, $\Gamma$, $\Gamma_{\mathrm{n}}$ and $\bar{\Gamma}_{\mathrm{n}}$ are penalty factors, which are sequentially increased to guarantee convergence.
\begin{remark}
When convergence is reached, i.e., the difference between the optimization variables and their previous optimal variables goes to zero, we have $T_{\mathrm{E}}= \mathrm{Prob}(R_{\mathrm{E}}\leq R_{\mathrm{E}}^{\mathrm{max}})$ and $T_{\mathrm{D}}= \mathrm{Prob}(P_{\mathrm{D}}\leq P_{\mathrm{D}}^{\mathrm{max}})$. This means that in the case of convergence, \eqref{eq:prob-2:eve} and \eqref{eq:prob-2:emf} can only be hold if the original constraints \eqref{eq:prob-1:eve} and \eqref{eq:prob-1:emf} are fulfilled. Hence, the fulfilment of the original constraints is guaranteed for the case of convergence.
\end{remark}
\begin{remark}
The structure of the reformulated problem also allows to set specific variables to zero and thus analyze special cases such as the case where artificial noise is applied by only one or none of the devices.
\end{remark}

\section{Numerical Results}

To evaluate the system performance numerically, we consider a setup where the covariance matrices of the eavesdropper channels and the channels in the exposure constraint are both $\bm{G}_{\mathrm{E}}=\bm{G}_{\mathrm{D}}=\bm{I}_{N}$ and $\bm{\bar{G}}_{\mathrm{E}}=\bm{\bar{G}}_{\mathrm{D}}=\bm{I}_{\bar{N}_{\mathrm{T}}}$. The channel from the BS to the UE is Gaussian distributed as $\bm{H}_{\mathrm{U}}^T\sim\mathcal{CN}(\bm{0},\bm{I}_{N})$, while the estimate of the self-interference channel is described by $\bar{G}_{\mathrm{U}}^{\mathrm{max}}=0.1$. The covariance matrices of the AWGN are $\bm{V}_{\mathrm{U}}=0.1\bm{I}_{\bar{N}_{\mathrm{R}}}$ and $V_{\mathrm{E}}=0.1$. Further, the transmit power is limited by $P^{\mathrm{max}}=\bar{P}^{\mathrm{max}}=\SI{10}{\dB}$; $\varepsilon$ and $\delta$ are both $0.05$.
All plots show the averaged results over $90$ realizations.
\begin{figure}[t]
    \centering
    \includegraphics{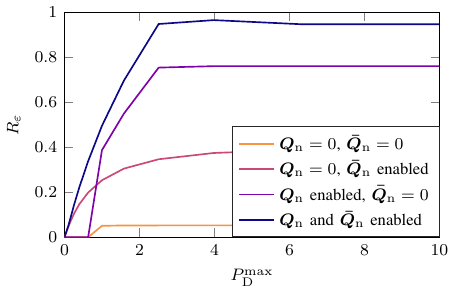}
    \caption{Secrecy rate over $P_{\mathrm{D}}^{\mathrm{max}}$ for different combinations of noise.}
    \label{fig:r_secrecy_avg}
\end{figure}

\begin{figure}
    \centering
    \subfloat[$\bm{Q}_{\mathrm{n}}=\bm{0}$, $\bm{\bar{Q}}_{\mathrm{n}}=\bm{0}$]{\includegraphics{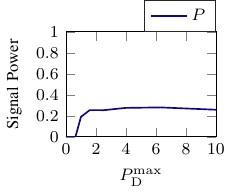}}\hfill
    \subfloat[$\bm{Q}_{\mathrm{n}}=\bm{0}$, $\bm{\bar{Q}}_{\mathrm{n}}$ enabled]{\includegraphics{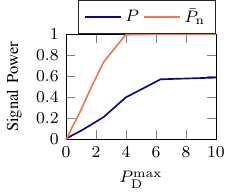}}\\
    \subfloat[$\bm{Q}_{\mathrm{n}}$ enabled, $\bm{\bar{Q}}_{\mathrm{n}}=\bm{0}$]{\includegraphics{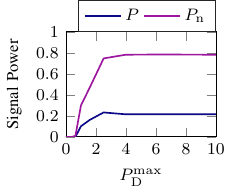}}\hfill
    \subfloat[$\bm{Q}_{\mathrm{n}}$ and $\bm{\bar{Q}}_{\mathrm{n}}$ enabled \label{fig:both_noise}]{\includegraphics{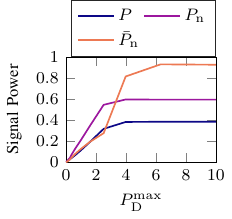}}
    \caption{Average allocated transmit power over $P_{\mathrm{D}}^{\mathrm{max}}$}
    \label{fig:power}
\end{figure}

For this setup, we first simulate the secrecy rate for all combinations of enabled/disabled artificial noise over the exposure limit $P_{\mathrm{D}}^{\mathrm{max}}$, and show the results in \figurename~\ref{fig:r_secrecy_avg}. The diagram illustrates that the secrecy rate enhances significantly when artificial noise becomes enabled. Especially for a large value of $P_{\mathrm{D}}^{\mathrm{max}}$, the impact of the artificial noise transmission from the BS is larger than the effect of the artificial noise transmission from the UE. The reason for this is that zero-forcing the artificial noise from the BS is possible, such that the interference at the UE vanishes. Compared to this, the artificial noise from the UE always causes a residual self-interference. With a more strict exposure constraint, the secrecy rate decreases. Below a certain value of $P_{\mathrm{D}}^{\mathrm{max}}$, the exclusive use of artificial noise from the BS can lead to feasibility issues. Hence, in this case, artificial noise from the UE becomes the more crucial noise source. 

To obtain further insights on the contribution of the optimized beamforming strategies, we show the allocated transmit power for the different beamformers in \figurename~\ref{fig:power}. Within the legends of the four plots, we employ the definitions, $P=\mathrm{trace}\{\bm{Q}\}$, $P_{\mathrm{n}}=\mathrm{trace}\{\bm{Q}_{\mathrm{n}}\}$ and $\bar{P}_{\mathrm{n}}=\mathrm{trace}\{\bm{\bar{Q}}_{\mathrm{n}}\}$. The results show that for a small $P_{\mathrm{D}}^{\mathrm{max}}$, the allocated transmit powers increase almost linearly as long as this constraint is active, before the allocated transmit power becomes constant. When both types of artificial noise are enabled (\figurename~\ref{fig:both_noise}), the power constraint at the BS is reached before the power constraint at the UE, such that the power of the artificial noise from the UE further increases for intermediate $P_{\mathrm{D}}^{\mathrm{max}}$.

\section{Conclusion}
In this paper, we have provided exact closed-form expressions of the SOP for MISO eavesdropper channels with Rayleigh fading and the probability that a specific exposure level is exceeded for the considered setup. Based on these expressions, an algorithm was proposed to optimize the $\varepsilon$-outage secrecy rate iteratively. Numerical results show that the transmission of artificial noise from BS and UE can both enhance the $\varepsilon$-outage secrecy rate. However, for a moderate exposure constraint, artificial noise transmission from the BS has a larger impact than artificial noise transmissions from the UE as it can be zero-forced, while always a residual self-interference of the artificial noise transmitted by the UE remains. Furthermore, for a strict exposure constraint, artificial noise from the UE has the larger impact.

\appendices
\section{Parameters of the Taylor Approximations}\label{appendix:taylor_parameters}

In \eqref{eq:eve:general:convex}, $C_{\mathrm{E},0}^\star$ is the value of \eqref{eq:prob:modified} in the operation point. The variables $\bm{C}_{\mathrm{E}}$, $\bm{C}_{\mathrm{E},\mathrm{n}}$, $\bm{\bar{C}}_{\mathrm{E},\mathrm{n}}$ and $C_{\mathrm{E},\mathrm{R}}$ represent the derivatives of \eqref{eq:prob:modified} towards the optimization variables $\bm{Q}$, $\bm{Q}_{\mathrm{n}}$, $\bm{\bar{Q}}_{\mathrm{n}}$ and $R_{\mathrm{E}}^{\mathrm{max}}$ in the operation point. To obtain these expressions, first the eigenvalue decomposition of $2\bm{L}_{\mathrm{E}}\left(\bm{Q}^\star-\left(2^{R_{\mathrm{E}}^{\mathrm{max},\star}}-1\right)\bm{Q}_{\mathrm{n}}^\star\right)\bm{L}_{\mathrm{E}}^H$ and $-2\left(2^{R_{\mathrm{E}}^{\mathrm{max},\star}}-1\right)\bm{\bar{L}}_{\mathrm{E}}\bm{\bar{Q}}_{\mathrm{n}}^\star\bm{\bar{L}}_{\mathrm{E}}^H$ needs to be defined as $\bm{U}_{\mathrm{E}}^\star\bm{\Lambda}_{\mathrm{E}}^\star\left(\bm{U}_{\mathrm{E}}^\star\right)^H$ and $\bm{\bar{U}}_{\mathrm{E}}^\star\bm{\bar{\Lambda}}_{\mathrm{E}}^\star\left(\bm{\bar{U}}_{\mathrm{E}}^\star\right)^H$, respectively. Based on this, the values of $\bm{C}_{\mathrm{E}}$, $\bm{C}_{\mathrm{E},\mathrm{n}}$, $\bm{\bar{C}}_{\mathrm{E},\mathrm{n}}$ and $C_{\mathrm{E},\mathrm{R}}$ can be obtained as
\begin{align}
    \bm{C}_{\mathrm{E}}&=2\bm{L}_{\mathrm{E}}^H\bm{U}_{\mathrm{E}}^\star\bm{F}_{\mathrm{E}}\left(\bm{U}_{\mathrm{E}}^\star\right)^H\bm{L}_{\mathrm{E}},\\
    \bm{C}_{\mathrm{E},\mathrm{n}}&=-2\left(2^{R_{\mathrm{E}}^{\mathrm{max},\star}}-1\right)\bm{L}_{\mathrm{E}}^H\bm{U}_{\mathrm{E}}^\star\bm{F}_{\mathrm{E}}\left(\bm{U}_{\mathrm{E}}^\star\right)^H\bm{L}_{\mathrm{E}},\end{align}\begin{align}
    \bm{\bar{C}}_{\mathrm{E},\mathrm{n}}&=-2\left(2^{R_{\mathrm{E}}^{\mathrm{max},\star}}-1\right)\bm{\bar{L}}_{\mathrm{E}}^H\bm{\bar{U}}_{\mathrm{E}}^\star\bm{\bar{F}}_{\mathrm{E}}\left(\bm{\bar{U}}_{\mathrm{E}}^\star\right)^H\bm{\bar{L}}_{\mathrm{E}},\\
    C_{\mathrm{E},\mathrm{R}}&=-2\ln(2)2^{R_{\mathrm{E}}^{\mathrm{max},\star}}\Big(\mathrm{trace}\left(\left(\bm{U}_{\mathrm{E}}^\star\right)^H\bm{L}_{\mathrm{E}}\bm{Q}_{\mathrm{n}}^\star\bm{L}_{\mathrm{E}}^H\bm{U}_{\mathrm{E}}^\star\bm{F}_{\mathrm{E}}\right)\nonumber\\&\hspace{-0.6cm}+\mathrm{trace}\left(\left(\bm{\bar{U}}_{\mathrm{E}}^\star\right)^H\bm{\bar{L}}_{\mathrm{E}}\bm{\bar{Q}}_{\mathrm{n}}^\star\bm{\bar{L}}_{\mathrm{E}}^H\bm{\bar{U}}_{\mathrm{E}}^\star\bm{\bar{F}}_{\mathrm{E}}\right)\Big)+\ln(2)2^{R_{\mathrm{E}}^{\mathrm{max}}}V_{\mathrm{E}}\times\nonumber\\&\hspace{-0.6cm}\sum_{\{k|\lambda_k > 0\}}\left(\frac{z_{\mathrm{E}}}{\lambda_k}-\Upsilon_k\right)^{m_k-1}\frac{\exp\left(-\lambda_k^{-1}z_{\mathrm{E}}\right)\left(\lambda_k\right)^{\sum_{j\neq k}m_j-1}}{(m_k-1)!\prod_{j\neq k}\left(\lambda_k-\lambda_j\right)^{m_j}}.
\end{align}
Thereby, $\bm{F}_{\mathrm{E}}$ and $\bm{\bar{F}}_{\mathrm{E}}$ are diagonal matrices, where each diagonal element $b_{\mathrm{E},l}$ and $\bar{b}_{\mathrm{E},l}$ represents the derivatives of \eqref{eq:prob:modified} by the corresponding $\lambda_k$ multiplied by $1/m_k$.
The diagonal entries of $\bm{\Lambda}_{\mathrm{E}}$ are arbitrary real values, while the diagonal entries of $\bm{\bar{\Lambda}}_{\mathrm{E}}$ are always negative or zero. 
In the case that a diagonal entry is negative or zero, the derivative (and $m_k$ times the elements of $\bm{F}_{\mathrm{E}}$ and $\bm{\bar{F}}_{\mathrm{E}}$) can be obtained as
\begin{align}
    F_k^-&=-\sum_{\{k'|\lambda_{k'} > 0,k'\neq k\}}\Bigg(\exp\left(-\Upsilon_{k'}\right)Q\left(m_{k'},\frac{z_{\mathrm{E}}}{\lambda_{k'}}-\Upsilon_{k'}\right)\nonumber\\&\times\frac{m_k\lambda_k}{\left(\lambda_{k'}-\lambda_k\right)^2}-\frac{\exp\left(-\frac{-z_{\mathrm{E}}}{\lambda_{k'}}\right)}{(m_{k'}-1)!}\left(\frac{z_{\mathrm{E}}}{\lambda_{k'}}-\Upsilon_{k'}\right)^{m_{k'}-1}\nonumber\\&\times\frac{m_k\lambda_{k'}}{\left(\lambda_{k'}-\lambda_k\right)^2}\Bigg)\prod_{j\neq {k'}}\left(\frac{\lambda_{k'}}{\lambda_{k'}-\lambda_j}\right)^{m_j}.
\end{align}
In the case that entries of $\bm{\Lambda}_{\mathrm{E}}$ are positive, the derivative (and thus $m_k$ times the according element of $\bm{F}_{\mathrm{E}}$) equals
\begin{align}
F_k^+&=\Bigg(-\exp\left(-\Upsilon_k\right)Q\left(m_k,\frac{z_{\mathrm{E}}}{\lambda_k}-\Upsilon_k\right)\sum_{j\neq k}\frac{m_j\lambda_j^2}{\lambda_k\left(\lambda_k-\lambda_j\right)^2}\nonumber\\&\nonumber+\frac{\exp\left(-\frac{z_{\mathrm{E}}}{\lambda_k}\right)}{\left(m_k-1\right)!}\left(\frac{z_{\mathrm{E}}}{\lambda_k}-\Upsilon_k\right)^{m_k-1}\nonumber\\&\left(-\frac{z_{\mathrm{E}}}{\lambda_k^2}+\sum_{j\neq k}\frac{m_j\lambda_j}{\left(\lambda_k-\lambda_j\right)^2}\right)\Bigg)\prod_{j\neq k}\left(\frac{\lambda_k}{\lambda_k-\lambda_j}\right)^{m_j}+F_k^-.
\end{align}

In \eqref{eq:emf:general:convex}, $C_{\mathrm{D},0}^\star$ is the value of $\mathrm{Prob}(P_{\mathrm{D}}\leq P_{\mathrm{D}}^{\mathrm{max}})$ in the operation point, while $\bm{C}_{\mathrm{D}}$ and $\bm{\bar{C}}_{\mathrm{D}}$ are, similar to $\bm{C}_{\mathrm{E}}$, defined as
\begin{align}
    \bm{C}_{\mathrm{D}}&=2\bm{L}_{\mathrm{D}}^H\bm{U}_{\mathrm{D}}^\star\bm{F}_{\mathrm{D}}\left(\bm{U}_{\mathrm{D}}^\star\right)^H\bm{L}_{\mathrm{D}},\\
    \bm{\bar{C}}_{\mathrm{D}}&=2\bm{\bar{L}}_{\mathrm{D}}^H\bm{\bar{U}}_{\mathrm{D}}^\star\bm{\bar{F}}_{\mathrm{D}}\left(\bm{\bar{U}}_{\mathrm{D}}^\star\right)^H\bm{\bar{L}}_{\mathrm{D}}.
\end{align}
Within these terms, $\bm{F}_{\mathrm{D}}$ and $\bm{\bar{F}}_{\mathrm{D}}$ are defined similar to $\bm{F}_{\mathrm{E}}$, while $\bm{U}_{\mathrm{D}}^\star$ and $\bm{\bar{U}}_{\mathrm{D}}^\star$ are the eigenvector matrices of $2\bm{L}_{\mathrm{D}}\left(\bm{Q}^\star+\bm{Q}_{\mathrm{n}}^\star\right)\bm{L}_{\mathrm{D}}^H$ and $2\bm{\bar{L}}_{\mathrm{D}}\bm{\bar{Q}}_{\mathrm{n}}^\star\bm{\bar{L}}_{\mathrm{D}}^H$, respectively.

\bibliographystyle{IEEEtran}
\bibliography{main}
\end{document}